\documentclass[11pt]{article}
\usepackage{
graphicx,amsmath,amssymb,bm}
\begin{document}

\title{ The second law of thermodynamics\\ from concavity of  energy eigenvalues}  

\author{
C. Itoi$^1$
and M. Amano$^2$\\
$^1$Department of Physics,  GS $\&$ CST,
Nihon University,\\
$^2$Department of Applied Physics, School of Engineering, The University of Tokyo, }

\maketitle

\abstract{Quantum dynamics controlled  by a time-dependent coupling constant are studied. 
 It is  proven that an energy eigenstate expectation value of work done by the 
 system in a quench process cannot exceed the work in the corresponding quasi-static process, 
 if and only if the energy eigenvalue is a concave function of the coupling constant. 
 We propose this concavity of energy eigenvalues as a new universal criterion for quantum dynamical systems 
 to  satisfy the second law of thermodynamics.
 We argue  simple universal conditions on quantum systems for the concavity, and show 
that every energy eigenvalue is indeed 
concave in some specific quantum systems.  These results agree with  the maximal work principle for adiabatic quench and quasi-static processes
 as an expression of the second law of thermodynamics.  Our result gives a simple example of an integrable system satisfying  an analogue to
 the strong  eigenstate thermalization hypothesis (ETH)  with respect to the principle of maximum work.}

\maketitle

\section{Introduction}
Prove the second law of thermodynamics within the framework of mechanics. 
This fundamental problem has  attracted  many physicists for more than hundred years. 
Many  attempts to clarify  the second law of thermodynamics
 in classical or quantum  mechanics
have been made. 
Roughly speaking, there are two categories of these studies. 
One is to clarify the  thermalization in isolated mechanical systems, 
and another one is to verify the impossibility to make any perpetual motion of second kind.    
In the first category,  a remarkable philosophy 
of typical pure quantum state for  thermally equilibrium states  has been proposed recently.
This  states that 
a typical pure state  in all possible mechanical states can be a thermally equilibrium state, and
any state  approaches to a typical state belonging to a set of all states most likely \cite{T0,GHT,T,GLMTZ,SS}. 
The eigenstate thermalization hypothesis (ETH) assumes that  almost all energy eigenstates are thermally equilibrium states \cite{B,D,LR,S,RDO,PSSV,DKPR,WGRE}.
The strong version  of ETH claims that all energy eigenstates are  thermally equilibrium states \cite{RDO,JS,SHP,SKNGG,BMH,KIH,KSG,MFSR,GG,DLL,YIS}.
There are several studies of ETH for integrable and non-integrable systems, and some of these have shown that the strong ETH fails for integrable systems 
\cite{RDO,JS,SHP,YIS,R,BKL,A,EF,VR}.  
 Interactions among many particles are essential to understand thermalization. 
 
In another category,  there have been  fewer studies than those in the first category. 
 Planck's  principle  is well-known as a specific expression of the impossibility  of perpetual motion of second kind.    
This principle claims  that a positive amount of work cannot be extracted  from  
thermodynamic systems in an arbitrary cyclic operation in the adiabatic environment.     
 Any realization of a perpetual motion of second kind implies the violation of this principle.  
 Passivity is known as  a sufficient condition on a quantum state  for the impossibility of work extraction from quantum systems.  
We say that a quantum state is passive, if a positive amount of work cannot be extracted from this state as an initial state 
in any cyclic unitary evolution. 
Some mixed states, such as the Gibbs states are passive  \cite{PW,L}, but general pure states except the ground state are not passive.
The passivity on quantum states is  stronger than Planck's principle, 
since a set of all unitary  transformations contains unphysical time evolution. Therefore, it is worth studying possibility or 
 impossibility of work extraction for an arbitrary  cyclic process given by a physical evolution from an initial pure sate.
 There have been several studies on the work extraction by a physical evolution  from an initial  pure state   
  in an adiabatic environment \cite{DCHFGV,GRE,PRGWE,MR,LLMPP,JSDMCG,SKCSG,DPZ,KIS}.  
 From the view point of the second law of thermodynamics,  however,
 a few studies of the work extraction have been reported.  
 Kaneko, Iyoda and Sagawa  study the validity of Planck's principle analogue to the strong ETH  in quantum spin systems  \cite{KIS}.
 They obtain  numerical evaluations of  amounts of works from  energy eigenstates of quantum spin systems in a composite cyclic process 
 which consists of  instantaneous changes of the interaction 
 and leaving the system a while.  They find the strong ETH like behavior  only in non-integrable systems.
 In integrable systems, however,  they find only weak ETH behavior.     
 In a slightly different problem setting from the work extraction,  
Mori studies irreversibility of  quench processes in quantum systems in adiabatic environment \cite{M}. 
Under an assumption of  ETH, he has proven 
an increasing entropy of quantum systems by an arbitrary adiabatic quench process.  
 
 In the present paper,  we propose a new universal criterion ``concavity of energy eigenvalues"
  for quantum mechanical systems 
 to satisfy the principle of maximum work for adiabatic quench and quasi-static  processes 
 as an expression of the second law of thermodynamics. 
 This criterion is expected to be useful, since it can be checked in each concerned system within the framework of time-independent 
 quantum mechanics.  The principle of maximum work for adiabatic processes states that work done by the 
 system in an arbitrary process cannot exceed the work in the corresponding quasi-static process.
This  is equivalent to Planck's principle, since the quasi-static process is reversible. 
We study quantum dynamical processes controlled by a time-dependent coupling constant.  
We evaluate  the changes of the energy under quench and quasi-static processes 
 solving the time-dependent  Schr\"odinger equation and using the adiabatic theorem.
We derive several formulae among expectation values of important physical quantities 
in an arbitrary energy eigenstate using the Hellmann-Feynman theorem \cite{H,F}. 
These formulae enable us to  prove that an energy eigenstate expectation value of 
work done by the system in a quench process cannot exceed the work in the corresponding quasi-static process, 
if and only if its energy eigenvalue is a concave function of the coupling constant. 
The proof of this irreversibility has been done by a purely quantum mechanical method  without any assumptions like ETH. 
 We give several simple examples in a specific many-particle systems confined in 
 a bounded region by a confining potential with a time-dependent  coupling constant
 where every  energy eigenvalue is a concave function of the coupling constant.  
  A system of free particles is an example of integrable system which satisfies an analogue to the strong ETH 
 for the principle of maximum work for adiabatic quench and quasi-static processes. 
 Finally, we point out that  a possible reason  for the fail of the strong  ETH like behavior of  integrable systems
reported in preceding studies is violation of the adiabatic theorem by the level crossing of the systems.

\section{Concavity of energy eigenvalues}
Here, we consider a general quantum many-body system. 
Let  $H$ and $H_I$ be self-adjoint operators and 
define a total Hamiltonian  ${\cal H}(\lambda)$ which consists of an unperturbed Hamiltonian 
 $H$ perturbed by  $ H_I $  with a coupling constant $\lambda \in {\mathbb R}$
\begin{eqnarray}
&&{\cal H}(\lambda):= H + \lambda H_I.  
\end{eqnarray}
To consider a time-dependent process of this quantum system,
define a function $\lambda: [t_0,t_1] \to {\mathbb R}
$ for $t_0<0<t_1$. 
Define a time-dependent Hamiltonian  ${\cal H}(\lambda(t))$ which consists of  a time-dependent term $\lambda(t) H_I $ 
and another  time-independent term  $H$  
\begin{eqnarray}
&&{\cal H}(\lambda(t)):= H + \lambda(t) H_I,  
\end{eqnarray}
where $\lambda(t)$ is regarded as a time-dependent coupling constant.
Assume that the ground state in ${\cal H} (\lambda)$ exists for any $\lambda$.  
The time-dependent state $\Psi(t)$ satisfies the following Schr\"odinger equation
\begin{equation}
i \hbar \frac{\partial }{\partial t} \Psi( t) = {\cal H}(\lambda(t)) \Psi(t).
\label{Schr}
\end{equation}
 Consider a process  of the quantum system from an initial state $\Psi(0)$
developed by the Schr\"odinger equation (\ref{Schr}). 
The amount of  work $W[\lambda]$ done by the quantum system in this process controlled by $\lambda(t)$  is defined by
\begin{equation}
W[\lambda, \Psi(0)]=- (\Psi(t_1), {\cal H}(\lambda(t_1) \Psi(t_1) ) +  (\Psi(0), {\cal H}(\lambda(0) )\Psi(0) ).
\end{equation}
Here, we consider two extremum of quasi-static (anneal) and quench  processes. 
To define the amount of work in a quasi-static (anneal) process, define a time-dependent coupling constant 
$\lambda_{t_1}: [t_0,t_1] \to {\mathbb R}$ for a given $\lambda_0, \lambda_1 >0$
 for arbitrary  $t_0 < 0 < t_1$, such that  $\lambda_{t_1}(t)  = \lambda_0 $  for $t \in [t_0, 0)$,
  $\lambda_{t_1}(t_1)  = \lambda_1$ and  $|\dot \lambda (t)| \leq C t_1^{-1}$ 
     for some $C>0$.
The amount of work in the quasi-static (anneal) process is defined by 
\begin{equation}
W_a[\Psi(0)]:= \lim_{t_1\to \infty}W[ \lambda_{t_1}, \Psi(0)]. 
\end{equation}
To define the amount of work in a quench process,  define $\lambda_q: [t_0,t_1] \to {\mathbb R}$ for a given $\lambda_0, \lambda_1 >0$ for an arbitrary $t_0 < 0 < t_1$
by $\lambda_q(t)  = \lambda_0 $ for $t \in [t_0, 0)$
and $\lambda_q(t)  = \lambda_1$ for $t \in [0, t_1]$.
The amount of work in the quanch process is defined by 
\begin{equation}
W_q[\Psi (0)]:= W[ \lambda_q, \Psi(0)]. 
\end{equation}
To examine ETH-like behavior of quantum system in these two processes, consider an energy eigenstate as an initial state. 
Let $\Phi_k(\lambda)$ be a normalized energy eigenstate which  belongs to the eigenvalue ${\cal E}_k(\lambda)$ as a function of $\lambda$
\begin{equation}
 {\cal H}(\lambda) \Phi_k = {\cal E}_k(\lambda) \Phi_k(\lambda).
\end{equation}
The following theorem gives necessary and sufficient condition on  energy eigenvalues of the quantum system for 
the validity of the principle of  the maximum work  in the quasi-static(anneal)  and quench processes.\\

\noindent
{\bf Theorem }{\it Consider  a  quantum system where the adiabatic theorem is valid, and 
consider the quasi-static  (anneal) and quench  processes  from an initial state  $\Phi_k(\lambda_0)$ of the quantum system
with an initial coupling constant $\lambda_0$ and  a final one $\lambda_1$
defined by the above. 
The amount of the work $W_q[\Phi_k(\lambda_0)]$ in the quench process 
cannot exceed the work $W_a[\Phi_k(\lambda_0)]$ in the corresponding quasi-static process
$$
W_q[\Phi_k(\lambda_0)] \leq W_a[\Phi_k(\lambda_0)],
$$
for arbitrary $\lambda_0, \lambda_1$,  
if and only if ${\cal E}_k(\lambda)$ is concave function of
the coupling constant $\lambda$. \\

\noindent
Proof. }
Let us  evaluate quench work first.
 
For $t<0$, let $$\Psi(t)= \exp[ -i t {\cal E}_k(\lambda_0)/\hbar]\Phi_k(\lambda_0)$$ be an initial state, and
consider a process by a quench $\lambda_q$ given by $\lambda_0\to\lambda_1$ at $t=0$ .   
The continuity of the state at $t=0$ implies that the state after the quench  is given by
\begin{equation}
\Psi( t) = \exp\Big[-it {\cal H}(\lambda_1) /\hbar \Big]\Phi_k(\lambda_0),
\end{equation}
The work done by the system in this quench process is 
\begin{eqnarray}
&&W_q={\cal E}_k(\lambda_0)-\langle {\cal H}(\lambda_1)\rangle_{\Psi} \nonumber \\
&&=\langle (H+\lambda_0 H_I)\rangle_{\Phi_k(\lambda_0)}-\langle (H+\lambda_1 H_I)\rangle_{\Phi_k(\lambda_0)} \nonumber \\
&&=(\lambda_0- \lambda_1 )\langle H_I
\rangle_{\Phi_k(\lambda_0)}.
\label{quenchwork}
\end{eqnarray}
The derivative of the eigenvalue equation with respect to the coupling constant
\begin{equation}\frac{\partial }{\partial \lambda}  \Big[H + \lambda H_I \Big] 
\Phi_k(\lambda) = \frac{\partial }{\partial \lambda} {\cal E}_k(\lambda) \Phi_k(\lambda), \end{equation}  
 gives
\begin{eqnarray}
&&\Big[H + \lambda  H_I \Big] \frac{\partial \Phi_k}{\partial \lambda} + H_I\Phi_k(\lambda) \\
&&= {\cal E}'_k(\lambda) \Phi_k+ {\cal E}_k(\lambda) \frac{\partial \Phi_k}{\partial \lambda}.
 \end{eqnarray}
The internal products between $\Phi_k (\lambda)$ and both sides give
\begin{equation}
\langle H_I \rangle_{\Phi_k(\lambda)} = {\cal E}_k'(\lambda).
\label{qE'}
\end{equation}
since the Hamiltonian ${\cal H}(\lambda)$ is self-adjoint. 
This identity is known as the Hellmann-Feynman theorem \cite{H,F}. 
The amount of work in the quench process has the following representation
\begin{equation} 
W_q[\Phi_k(\lambda_0)]=(\lambda_0-\lambda_1)\langle H_I \rangle_{\Phi_k(\lambda)}= (\lambda_0-\lambda_1) {\cal E}_k'(\lambda_0).
\end{equation}
Next, let us evaluate amount of work in the corresponding quasi-static (anneal process $\lambda_0\to\lambda_1$. 
Under an assumption of the  the quantum adiabatic theorem,
the work done by the system in the quasi-static process is given by
$$
W_a[\Phi_k(\lambda_0)]={\cal E}_k(\lambda_0)-{\cal E}_k(\lambda_1),
$$
since a quasi-static process preserves a quantum number of the energy
eigenstate.

 The following is obtained.  
 If and only if  ${\cal E}_k(\lambda)$ is a concave function of $\lambda$, 
 \begin{equation}
(\lambda_0-\lambda_1) {\cal E}_k'(\lambda_0)\leq {\cal E}_k(\lambda_0) -{\cal E}_k(\lambda_1).
 \end{equation}    
is valid for  any $\lambda_0, \lambda_1$. 
This implies that  the work $W_q[\Phi_k(\lambda_0)]$ in the quench process from the initial energy eigenstate $\Phi_k$
 cannot exceed the work  $W_a[\Phi_k(\lambda_0)]$ in the corresponding  quasi-static process  for any $\lambda_0\to\lambda_1$.
 This completes the proof. $\Box$\\

Here, we argue some nature of  concavity. 
We have the  following formula for quantum systems
\begin{equation}
 {\cal E}_k'(\lambda) =\frac{{\cal E}_k(\lambda)-\langle H \rangle_{\Phi_k(\lambda)}}{\lambda}.
 \label{qE'2}
\end{equation}
This and (\ref{qE'})  give another one 
\begin{equation}
 {\cal E}_k''(\lambda) =-\frac{1}{\lambda} \frac{d \langle H \rangle_{\Phi_k(\lambda) }}{d \lambda}. 
 \label{qE''2}
\end{equation}
For the case $\lambda > 0$,    the expression (\ref{qE''2}) and the concavity ${\cal E}''_k(\lambda) \leq 0$ require that  
$\langle H \rangle_{\Phi_k(\lambda)}$ is a monotonically increasing function of $\lambda$. 
For the case $\lambda < 0$,  
the expression (\ref{qE''2}) and the concavity   ${\cal E}_k''(\lambda) \leq 0$ require that  $\langle H \rangle_{\Phi_k(\lambda)}$ 
is  monotonically decreasing function of   $\lambda$.

Addition to the above argument on the monotonically increasing function of the expectation value of the unperturbed Hamiltonian
for the concavity of ${\cal E}_k(\lambda)$, we give another argument in a perturbative calculation.
Let
$$\Phi_k(\lambda+\Delta \lambda) \simeq \Phi_k(\lambda) + \Delta \lambda  \Phi_k^{(1)}(\lambda)+ \Delta \lambda^2  \Phi_k^{(2)}(\lambda)
+ \cdots .
$$ 
be a perturbation expansion of a  perturbed eigenstate at a coupling constant $\lambda + \Delta \lambda$. 
It is well-known that a suitable choice of 
the perturbation Hamiltonian $\Delta \lambda H_I$  leads to
 the orthogonality between the  first order state and the unperturbed state
 \begin{equation} 
 ( \Phi_k^{(1)}  (\lambda), \Phi_k (\lambda) )=0.
 \label{pertorth}
 \end{equation} 
In this case,  the first order correction is   given by
 $$ 
  \Phi_k^{(1)}(\lambda)=\sum_{l \neq k} \frac{(\Phi_k(\lambda) , H_I \Phi_l(\lambda)  )  }{{\cal E}_k(\lambda) -{\cal E}_l(\lambda)} \Phi_l(\lambda). 
 $$
The second order approximation of the perturbed energy eigenvalue at a coupling constant $\lambda + \Delta \lambda$
is given by  
\begin{eqnarray}
{\cal E}_k(\lambda+\Delta \lambda) 
&\simeq&  {\cal E}_k(\lambda) + \Delta \lambda (\Phi_k(\lambda) , H_I \Phi_k(\lambda)  ) 
 \\&+&\Delta \lambda^2 \sum_{l \neq k} \frac{|(\Phi_k(\lambda) , H_I \Phi_l(\lambda)  ) |^2 }{{\cal E}_k(\lambda) -{\cal E}_l(\lambda)} + \cdots .
\end{eqnarray}
The second and third terms in the  right hand side give the derivatives of energy eigenvalue
\begin{eqnarray}{\cal E}_k'(\lambda) &=& (\Phi_k(\lambda) , H_I \Phi_k(\lambda)  ),  \\
{\cal E}_k''(\lambda) &=&  2 \sum_{l \neq k} \frac{|(\Phi_k(\lambda) , H_I \Phi_l(\lambda)  ) |^2 }{{\cal E}_k(\lambda) -{\cal E}_l(\lambda)} 
\\&=&  2 \Big[ \sum_{{\cal E} _l < {\cal E}_k} \frac{|(\Phi_k(\lambda) , H_I \Phi_l(\lambda)  ) |^2 }{{\cal E}_k(\lambda) -{\cal E}_l(\lambda)} 
\\&-&\sum_{{\cal E} _l >{\cal E}_k} \frac{|(\Phi_k(\lambda) , H_I \Phi_l(\lambda)  ) |^2 }{{\cal E}_l(\lambda) -{\cal E}_k(\lambda)} \Big]
\end{eqnarray}
The first line reproduce the formula (\ref{qE'}) obtained by the Hellmann-Feynman theorem, and the second term gives the 
second derivative of the energy eigenvalue.   
If the finial line is negative semi-definite,  the eigenvalue is a concave function of $\lambda$. 
Generally speaking, the final line is expected to be negative semi-definite for an arbitrary low lying energy eigenstates with $k$  
as well as the ground state $k=0$, since number of energy eigenstates is increasing function of the 
energy eigenvalue. Only these low lying states can survive in the thermodynamic limit. 
Although this argument depends  on the matrix elements of $H_I$ and is not rigorous, non-positivity of ${\cal E}_k''(\lambda)$ is plausible in systems with many
degrees of freedom generally.

\section{Example}
Here we consider a quantum dynamical system of $N$ particles confined in a  bounded region 
 in one dimension. 
Let $x=(x_1, \cdots, x_N)$ be  a set of positions of $N$ particles, and
let $\Psi(x,t)$ be a time-dependent state  of particles satisfying the  Schr\"odinger equation.
\begin{equation}
i \hbar \frac{\partial }{\partial t} \Psi(x, t) = {\cal H}(\lambda(t)) \Psi(x,t).
\end{equation}
The time-dependent Hamiltonian ${\cal H}(\lambda(t))$ has a positive semi-definite 
confining potential $u(x_n)$
with a time-dependent coupling constant $\lambda(t)$ 
and another  time-independent term  $H$, which consists of
kinetic energy and interaction $V(x)$ among $N$ particles
\begin{eqnarray}
&&{\cal H}(\lambda):= H + \lambda H_I ,  \\
 &&H:=-\frac{\hbar^2}{2} \sum_{n=1} ^N \frac{\partial^2}{\partial x_n^2} + V(x),\\
&&H_I:= \sum_{n=1}^N u(x_n).
\end{eqnarray}
First, consider
the case $V(x)=0$.
Let  $\phi_{k_n} (x_n)$ be a single particle energy eigenstate of  $n$-th particle
with an eigenvalue  $E_{k_n}(\lambda)$. If this function  is concave function of $\lambda$ for any $n$, 
the theorem is valid also for any eigenstate of $N$ free particles for a set $k=(k_1, \cdots, k_N)$  of quantum numbers
$$
\Phi_k (x)=\prod_{n=1}^N   \phi_{k_n} (x_n),
$$
 with the eigenvalue
\begin{equation}
{\cal E}_{k}(\lambda)=
\sum_{n=1}^N E_{k_n} (\lambda).
\end{equation}  
Note that the energy of each particle is conserved for a fixed $\lambda$.  
This system is integrable, since it has $N$ coordinates
and $N$ conserved quantities. 
Let us give a simple example where all energy eigenvalues are concave functions of $\lambda$.  
Consider $N$ free particles in a potential  
\begin{equation}
u(x) =\frac{1}{2j} x^{2j},
\end{equation}
with a positive integer $j$. This potential $u(x)$ is positive semi-definite.
The virial  theorem \cite{Fock} and the identity (\ref{qE'}) imply
\begin{equation}
\langle H \rangle_{\Phi_k} =\sum_{n} \frac{\lambda}{2} \langle x_n u'(x_n) \rangle_{\Phi_k} =\lambda j\sum_{n=1}^N  \langle u(x_n) \rangle_{\Phi_k} 
=\lambda j  {\cal E}_k'(\lambda).
\end{equation}
This and the identity (\ref{qE'2}) give the following differential equation for ${\cal E}_k(\lambda)$
\begin{equation}
(j+1)\lambda {\cal E}_k'(\lambda) = {\cal E}_k(\lambda). 
\end{equation}
The solution of this differential equation is given by
 \begin{equation}
 {\cal E}_k(\lambda) = C_k \lambda^\frac{1}{j+1},
 \label{Elambda}
 \end{equation}
with a positive constant $C_k =O(N)$ which is extensive   and 
 independent of $\lambda$.  This shows the following  strict concavity of the function ${\cal E}_k(\lambda)$ 
$${\cal E}_k''(\lambda) =-\frac{j}{(j+1)^2} C_k \lambda^{-\frac{2j+1}{j+1}}  < 0$$ for any $k$.
Therefore,   the amount of work  in  the quench process   is strictly and extensively smaller   
 than  that in the quasi-static process  for every initial eigenstate $\Phi_k$. 
This result implies  an analogue to the strong ETH, despite the integrability of free particles.
The same argument based on the virial theorem is possible for a model with an interaction
\begin{equation}
\displaystyle V(x) := \sum_{1\leq m<n \leq N} \frac{g}{(x_m-x_n)^2}.
\end{equation}
In this system, we have the solution (\ref{Elambda}) and every energy eigenvalue is a concave function of $\lambda$.  
The generalization of our argument based on the virial theorem to  models in an arbitrary dimension is also possible.\\

Here, we explain  
ETH analogue for the principle of maximum work in the present paper and also
argued by Kaneko, Iyoda and Sagawa \cite{KIS}.   
The ETH claims that every energy eigenstate can be a thermal equilibrium state. 
The micro-canonical density operator $\rho({\cal E}) $ in an energy shell
$K({\cal E})=\{k \ ; \  {\cal E}<  {\cal E}_k(\lambda) <  {\cal E}+\Delta {\cal E}\}$
is defined by
$$
\rho ({\cal E}) :=Z^{-1} \sum_{k \in K({\cal E})
}  | \Phi_k(\lambda) \rangle \langle \Phi_k(\lambda) |,  
$$  
where  $Z$ is determined by a normalization ${\rm Tr} \rho ({\cal E} )=1$.  
Consider a density $X:= {\cal O}/N $ as a macro variable, where 
${\cal O} =\sum_{n=1}^N{\cal O}_n$ is an extensive variable with a bounded observable ${\cal O}_n$. 
Denote $x_{eq}= {\rm Tr} [ \rho({\cal E} ) X]$.  
The ETH defined by Tasaki \cite{T}
 states that for every energy eigenstate $\Phi_k$  in the energy shell  $k \in K({\cal E})$ 
  and for a fixed value $\delta > 0$, there exists $\gamma>0$ such that
$$( \Phi_k,  P_{|X-x_{eq} |> \delta} \Phi_k) \leq e^{-\gamma N},$$ 
 for any macro variable $X$, where $P_{|X-x_{eq}|>\delta}$ is projection operator onto the Hilbert space 
 spanned by eigenstates of $X$ with eigenvalues $x$ such that  $|x -x_{eq}| >  \delta.$      
 Therefore every energy eigenstate expectation value $( \Phi_k,  X \Phi_k)$ for $k \in K({\cal E})$
 is identical to the corresponding micro-canonical expectation value $x_{eq}$.
Mori has proven the irreversibility of a quench process of an arbitrary quantum system by
showing  the strict increasing  of a diagonal entropy  under the assumption of the ETH.   
We say that strong ETH is valid, if the ETH is valid for any energy shell.    
On the other hand, 
we say that strong  ETH analogue for the principle of maximum work is valid if  the amount of work done by a quantum system
 in an arbitrary adiabatic process $\lambda_0 \to \lambda_1$ from
every initial energy eigenstate cannot exceed  that  in the corresponding quasi-static adiabatic process $\lambda_0 \to \lambda_1$.
 In the present paper, we have proven the inequality for amounts of works in  quench and the quasi-static processes 
 from every  initial concave energy eigenstate without any thermodynamic assumptions.
The strong ETH analogue for the principle of maximum work  has been proven for an arbitrary quench process.

\section{Summary}
 We discuss quantum dynamics  governed  by a Hamiltonian with a time-dependent coupling constant.
We prove that the concavity of the energy eigenvalue in the coupling constant 
gives  necessary and  sufficient condition on the principle of maximum work for quench and quasi-static processes. This criterion can be checked in each concerned system within the frame work of time-independent quantum mechanics. 
We give an example with a many-particle system in a specific potential  
where every energy eigenvalue becomes a concave function of the coupling constant.    
This is an example of an integrable system which satisfies the strong ETH analogue with respect to Planck's principle.

Here, we comment on  
quantum systems with level crossing of energy eigenvalues by the change of coupling constant. 
In such systems, the adiabatic theorem fails, and concavity of the energy eigenvalues does not yields Planck's principle in strong sense
as pointed out by Kaneko, Iyoda and Sagawa \cite{KIS}. We consider that level crossing likely occurs in  integrable quantum systems. 
On the other hand in general non-integrable quantum systems,  the energy levels are repulsive and level crossing occurs  unlikely. 
Therefore,  the adiabatic theorem is expected to be valid in general quantum systems. \\

Acknowledgments     

We are grateful to R. M. Woloshyn for a careful reading of the manuscript.
It is pleasure to thank T. Sako and S. Suzuki for helpful discussions. \\

$^*$itoi@phys.cst.nihon-u.ac.jp

\end{document}